\documentclass[twocolumn]{aastex63}
\usepackage[utf8]{inputenc}
\usepackage{xspace}
\usepackage{xcolor}
\usepackage{url}
\usepackage{array}

\newcommand{\bjdtdb}{\ensuremath{\rm {BJD_{TDB}}}}
\newcommand{\feh}{\ensuremath{\left[{\rm Fe}/{\rm H}\right]}}

\newcommand{\teff}{\ensuremath{T_{\rm eff}}\xspace}
\newcommand{\logg}{\ensuremath{\log g}}

\newcommand{\msun}{\ensuremath{\,M_\Sun}}
\newcommand{\rsun}{\ensuremath{\,R_\Sun}}
\newcommand{\lsun}{\ensuremath{\,L_\Sun}}
\newcommand{\mj}{\ensuremath{\,M_{\rm J}}}
\newcommand{\rj}{\ensuremath{\,R_{\rm J}}}

\newcommand{\fave}{\langle F \rangle}
\newcommand{\fluxcgs}{10$^9$ erg s$^{-1}$ cm$^{-2}$}

\newcommand{\kms}{\,km\,s$^{-1}$}
\newcommand{\ms}{\,m\,s$^{-1}$}

\newcommand{\thisstar}{HD~118203\xspace}
\newcommand{\thisplanet}{HD~118203\,b\xspace}
\newcommand{\tess}{{\it TESS}\xspace}

\begin{document}

%\title{TESS Detects a Transit of an RV-Discovered Planet}
%\title{TESS Detects the Transit of the RV-Discovered Eccentric Planet HD 118203 b}
\title{TESS Reveals HD 118203 b to be a Transiting Planet}

%%%   Author list order is preliminary and arbitrary.  For now, please add your full
%%%   information at any location, and when we near a complete draft we will reorder based %%%   on contributions

\author[0000-0002-3827-8417]{Joshua Pepper}
\affiliation{Department of Physics, Lehigh University, 16 Memorial Drive East, Bethlehem, PA 18015, USA}

\author[0000-0002-7084-0529]{Stephen R.\ Kane}
\affiliation{Department of Earth and Planetary Sciences, University of California, Riverside, CA 92521, USA}

\author[0000-0001-8812-0565]{Joseph E.\ Rodriguez} %agreed
\affiliation{Center for Astrophysics \textbar \ Harvard \& Smithsonian, 60 Garden St, Cambridge, MA 02138, USA}
%\affiliation{Somewhere on a Tropical Jupiter}% HAHAHAHA -JR

\author[0000-0003-0595-5132]{Natalie R.\ Hinkel}
\affiliation{Southwest Research Institute, 6220 Culebra Rd, San Antonio, TX 78238, USA}

\author[0000-0003-3773-5142]{Jason D. Eastman} 
\affiliation{Center for Astrophysics \textbar \ Harvard \& Smithsonian, 60 Garden St, Cambridge, MA 02138, USA}

\author[0000-0002-6939-9211]{Tansu Daylan}
\affiliation{Department of Physics and Kavli Institute for Astrophysics and Space Research, Massachusetts Institute of Technology, Cambridge, MA 02139, USA}
\affiliation{Kavli Fellow}

\author[0000-0003-4603-556X]{Teo Mocnik}
\affiliation{Department of Earth and Planetary Sciences, University of California, Riverside, CA 92521, USA}

\author[0000-0002-4297-5506]{Paul A.\ Dalba}
\altaffiliation{NSF Astronomy and Astrophysics Postdoctoral Fellow}
\affiliation{Department of Earth and Planetary Sciences, University of California, Riverside, CA 92521, USA}

\author{Tara Fetherolf}
\affiliation{Department of Physics and Astronomy, University of California, Riverside, CA 92521, USA}

\author[0000-0002-3481-9052]{Keivan G.\ Stassun} 
\affil{Vanderbilt University, Department of Physics \& Astronomy, 6301 Stevenson Center Lane, Nashville, TN 37235, USA}

\author[0000-0002-4588-5389]{Tiago L.\ Campante}
\affiliation{Instituto de Astrof\'{\i}sica e Ci\^{e}ncias do Espa\c{c}o, Universidade do Porto,  Rua das Estrelas, 4150-762 Porto, Portugal}
\affiliation{Departamento de F\'{\i}sica e Astronomia, Faculdade de Ci\^{e}ncias da Universidade do Porto, Rua do Campo Alegre, s/n, 4169-007 Porto, Portugal}

\author[0000-0001-7246-5438]{Andrew Vanderburg}
\affiliation{Department of Astronomy, The University of Texas at Austin, Austin, TX 78712, USA}
\affiliation{NASA Sagan Fellow}

\author[0000-0001-8832-4488]{Daniel Huber}
\affiliation{Institute for Astronomy, University of Hawai`i, 2680 Woodlawn Drive, Honolulu, HI 96822, USA}

\author[0000-0003-0395-9869]{B.\ Scott Gaudi}
\affiliation{Department of Astronomy, The Ohio State University, 140 West 18th Avenue, Columbus, OH 43210, USA}

\author[0000-0002-9480-8400]{Diego Bossini}
\affiliation{Instituto de Astrof\'{\i}sica e Ci\^{e}ncias do Espa\c{c}o, Universidade do Porto,  Rua das Estrelas, 4150-762 Porto, Portugal}
% diego.bossini@astro.up.pt

\author[0000-0002-1835-1891]{Ian Crossfield}
\affiliation{Department of Physics and Kavli Institute for Astrophysics and Space Research, Massachusetts Institute of Technology, Cambridge, MA 02139, USA}

%  ----------------- TESS Architects --------------------

\author[0000-0003-2058-6662]{George~R.~Ricker}
\affiliation{Department of Physics and Kavli Institute for Astrophysics and Space Research, Massachusetts Institute of Technology, Cambridge, MA 02139, USA}
%  C

\author[0000-0001-6763-6562]{Roland~Vanderspek}
\affiliation{Department of Physics and Kavli Institute for Astrophysics and Space Research, Massachusetts Institute of Technology, Cambridge, MA 02139, USA}
%  C

\author[0000-0001-9911-7388]{David~W.~Latham}
\affiliation{Harvard-Smithsonian Center for Astrophysics, 60 Garden St, Cambridge, MA 02138, USA}

\author[0000-0002-6892-6948]{S.~Seager}
\affiliation{Department of Physics and Kavli Institute for Astrophysics and Space Research, Massachusetts Institute of Technology, Cambridge, MA 02139, USA}
\affiliation{Department of Earth, Atmospheric and Planetary Sciences, Massachusetts Institute of Technology, Cambridge, MA 02139, USA}
\affiliation{Department of Aeronautics and Astronautics, MIT, 77 Massachusetts Avenue, Cambridge, MA 02139, USA}
%  C

\author[0000-0002-4265-047X]{Joshua~N.~Winn}
\affiliation{Department of Astrophysical Sciences, Princeton University, 4 Ivy Lane, Princeton, NJ 08544, USA}
%  C

\author{Jon~M.~Jenkins}
\affiliation{NASA Ames Research Center, Moffett Field, CA, 94035, USA}
%  C

%  ----------------- SPOC Architects --------------------

\author[0000-0002-6778-7552]{Joseph~D.~Twicken}
\affiliation{NASA Ames Research Center, Moffett Field, CA, 94035, USA}
\affiliation{SETI Institute, Mountain View, CA 94043, USA}
% C     joseph.twicken@nasa.gov

\author[0000-0003-4724-745X]{Mark~Rose}
\affiliation{NASA Ames Research Center, Moffett Field, CA, 94035, USA}
% C     mark.rose@nasa.gov

\author[0000-0002-6148-7903]{Jeffrey C.\ Smith}
\affiliation{NASA Ames Research Center, Moffett Field, CA, 94035, USA}
\affiliation{SETI Institute, Mountain View, CA 94043, USA}
% C     jeffrey.c.smith-1@nasa.gov

%  ----------------- POC Architects --------------------

\author[0000-0002-5322-2315]{Ana~Glidden}
\affiliation{Department of Earth, Atmospheric and Planetary Sciences, Massachusetts Institute of Technology, Cambridge, MA 02139, USA}
\affiliation{Department of Physics and Kavli Institute for Astrophysics and Space Research, Massachusetts Institute of Technology, Cambridge, MA 02139, USA}
% C     aglidden@mit.edu

\author{Alan~M.~Levine}
\affiliation{Department of Physics and Kavli Institute for Astrophysics and Space Research, Massachusetts Institute of Technology, Cambridge, MA 02139, USA}
% C     aml@space.mit.edu

\author{Stephen Rinehart}
\affiliation{NASA Goddard Space Flight Center, Greenbelt, MD 20771}
% C     stephen.a.rinehart@nasa.gov

%  ----------------- SOC Architects --------------------

\author[0000-0001-6588-9574]{Karen~A.~Collins} 
\affiliation{Harvard-Smithsonian Center for Astrophysics, 60 Garden St, Cambridge, MA 02138, USA}
% C     karenacollins@outlook.com
 	
\author[0000-0003-3654-1602]{Andrew W. Mann}
\affiliation{Department of Physics and Astronomy, The University of North Carolina at Chapel Hill, Chapel Hill, NC 27599, USA} 
% C     mannaw@email.unc.edu

\author[0000-0002-0040-6815]{Jennifer~A.~Burt}
\affiliation{Jet Propulsion Laboratory, California Institute of Technology, 4800 Oak Grove drive, Pasadena CA 91109, USA}
% C     jennburt@mit.edu, jennifer.burt2@gmail.com

%  ----------------- KELT Architects --------------------

\author[0000-0001-5160-4486]{David~J.~James}
\affiliation{Center for Astrophysics $\vert$ Harvard \& Smithsonian, 60 Garden Street, Cambridge, MA 02138, USA}
\affiliation{Black Hole Initiative at Harvard University, 20 Garden Street, Cambridge, MA 02138, USA}

\author[0000-0001-5016-3359]{Robert~J.~Siverd}
\affiliation{Gemini Observatory, Northern Operations Center, 670 N. A’ohoku Place, Hilo, HI 96720, USA}

\date{\today}

\begin{abstract}

%Follow-up observations of known exoplanets continue to be an important component of exoplanetary science to provide characterization of orbits and atmospheres. 
The exoplanet \thisplanet, orbiting a bright ($V=8.05$) host star, was discovered using the radial velocity method by \citet{daSilva2006}, but was not previously known to transit. \tess photometry has revealed that this planet transits its host star. Five planetary transits were observed by \tess, allowing us to measure the radius of the planet to be $1.133^{+0.031}_{-0.030}R_J$, and to calculate the planet mass to be $2.173^{+0.077}_{-0.080}M_J$. The host star is slightly evolved with an effective temperature of $\teff = 5692\pm83$~K and a surface gravity of $\logg = 3.891^{+0.019}_{-0.020}$.  With an orbital period of $6.134980^{+0.000038}_{-0.000037}$ days and an eccentricity of $0.316\pm0.021$, the planet occupies a transitional regime between circularized hot Jupiters and more dynamically active planets at longer orbital periods. The host star is among the ten brightest known to have transiting giant planets, providing opportunities for both planetary atmospheric and asteroseismic studies.

\end{abstract}


%%%%%%%%%%%%%%%%%%%%%%%%%%%%%%%%%%%%%%%%%%%%%%%%%%%%%

\section{Introduction}
\label{sec:intro}

The dawn of planetary transit science began with the transit detection of planets that had been discovered with the radial velocity (RV) technique. The first of these was HD~209458~b \citep{Charbonneau2000,Henry2000} which, for several years thereafter, was the only known transiting planet. To date, there are eleven planets for which transits were detected after an initial RV discovery: HD~80606~b \citep{Fossey2009,Garcia-Melendo2009,Laughlin2009}, 55~Cancri~e \citep{Winn:2011,Demory:2011}, GJ~436~b \citep{Gillon:2007}, HD~149026~b \citep{Sato2005}, HD~189733~b \citep{Bouchy2005}, HD~17156~b \citep{Barbieri2007}, HD~97658~b \citep{Dragomir:2013}, GJ~3470~b \citep{Bonfils:2012}, HD~219134~b \citep{Motalebi:2015}, and HD~219134~c \citep{Gillon:2017}. These transiting exoplanets are important because their host stars are bright, especially relative to the typical hosts of transiting exoplanets
\citep{Kane2007,Kane2009}. The brighter host stars enable detailed follow-up observations to be carried out to study the planetary atmospheres, such as the {\it Spitzer} observations of HD~189733~b \citep{Knutson2007}. The Transit Ephemeris Refinement and Monitoring Survey (TERMS) has continued to observe known exoplanets using photometric and RV techniques to refine orbits and investigate a variety of stellar and planetary signatures \citep{Dragomir2011,Kane2011,Pilyavsky2011,Dragomir2012,Henry2012,Hinkel2015,Kane2016}. For the Transiting Exoplanet Survey Satellite (\tess), a primary goal is the detection of planets transiting bright host stars \citep{Ricker2015} and the subsequent characterization of the atmospheres of some of those planets \citep{Kempton2018}.

Through analysis of the \tess observation strategy and the known exoplanet demographics, \citet{Dalba2019} predicted that several known RV planets would be discovered to transit. The prediction was consistent with the early \tess discovery of an additional transiting planet in the $\pi$ Mensae system \citep{Huang2018}, which was already known to have a longer-period planet from earlier RV surveys. However, the large number of RV planets with orbital periods shorter than 10 days provides many opportunities to detect further transits among the known RV population \citep{Kane2008}. This paper reports the detection of transits of \thisplanet, a previously known giant planet ($M_p \sin i = 2.13$~$M_J$) in a 6.13-day eccentric ($e = 0.31$) orbit around a bright ($V = 8.05$) star \citep{daSilva2006}. The star was observed by \tess during Sector~15 of Cycle~2 observations of the northern ecliptic hemisphere. 

The stellar brightness of \thisstar combined with the eccentric nature of the orbit presents an important opportunity to study the atmospheres of exoplanets under tidal stress from the host star.  \thisplanet joins TOI-172~b \citep{Rodriguez:2019} and HD~2685~b \citep{Jones:2019} as \tess-detected giant planets in eccentric orbits close to their host stars. In Section~\ref{sec:obs} we describe the details of \tess observations and photometry, as well as ground-based observations that contribute to the analysis. Section~\ref{sec:system} presents a detailed analysis of the photometry and RVs in order to determine the system characteristics. In Section~\ref{sec:disc} we discuss the discovery within the context of the known exoplanet population and the prospects for further observations.

%%%%%%%%%%%%%%%%%%%%%%%%%%%%%%%%%%%%%%%%%%%%%%%%%%%%%

\section{\tess Observations}
\label{sec:obs}

The star \thisstar (see Table \ref{tab:LitProps} for additional names) was observed by \tess in Sector 15 of the mission.  The star was selected for 2-minute \tess observations for several reasons. It was included in the exoplanet candidate target list accompanying version 8 of the \tess Input Catalog of prime targets for \tess discovery of small exoplanets \citep{Stassun:2019} at a priority of 0.00282, placing it among the top 20\% of targets selected for transit detection. In addition, the target was proposed for observations by a number of guest investigators\footnote{G022197 (Shporer, A.), G022053 (Kane, S.), and G022103 (Huber, D.)}. Furthermore, \thisstar was included in the Asteroseismic Target List \citep{Schofield19} of solar-like oscillators to be observed in 2-min cadence with \tess. The Asteroseismic Target List comprises of bright, cool main-sequence and subgiant stars and forms part of the larger target list proposed by the \tess Asteroseismic Science Consortium. 

\tess obtained 17,839 observations of \thisstar in 2-minute cadence from 15 August to 10 September 2019.  As a 2-minute target, the light curve was processed by the SPOC data reduction pipeline \citep{Jenkins:2016}, and released through the \tess page of the MAST archive\footnote{https://archive.stsci.edu/tess/}.  The SPOC light curve clearly shows four complete transits, and one partial transit, at a period corresponding to the $\sim6.13$d known planet orbital period detected by \citet{daSilva2006}.  During the automated search for new planets, transits of \thisstar were identified by the Science Processing Operations Center (SPOC) transit search pipeline \citep{Jenkins:2002,Jenkins:2010}, and \thisstar was identified as a promising transit candidate by the \tess vetting procedure (Guerrero, et al.\ in preparation) using \tess data validation tools \citep{Twicken:2018,Li:2019} and assigned TOI number 1271.01.

The discovery paper by \citet{daSilva2006} provides 43 individual RV observations from the ELODIE spectrograph \citep{Baranne:1996}, acquired between May 2004 and July 2005.  That paper reports an orbital period for the companion of $P = 6.1335 \pm 0.0006$ days and an eccentricity of $e = 0.309 \pm 0.014$, with an additional acceleration of $49.7 \pm 5.7$ \ms yr$^{-1}$, and an average radial velocity of $-29.387 \pm 0.006$ \kms.  We have included those RV observations in our analysis below in \S \ref{sec:system}.

%%%%%%%%%%%%%%%%%%%%%%%%%%%%%%%%%%%%%%%%%%%%%%%%%%%%%

\section{System Analysis}
\label{sec:system}

In addition to the TESS photometry and the RV data and system parameters reported by \citet{daSilva2006}, we have gathered various properties of \thisstar from existing catalogs and archives.  These include elemental abundances, spectroscopic parameters, other measures of photometric variability, and kinematic information.  Those data and parameters are described below, in addition to the procedures we used to conduce a global fit of the system properties.

\subsection{Abundances and Effective Temperature}
\label{sec:abundances}

A total of 37 elements (including neutral and singly-ionized) were measured based on the spectrum of the host star \thisstar by 8 different groups \citep{Brugamyer11, Gonzalez10a, Gonzalez10b, Maldonado13, DelgadoMena15, Maldonado16, Luck17, Maldonado18}. The star is, in general, metal rich compared with the Sun, such that only a few elemental abundance ratios fall below solar ratios ([Li/H], [Cr II/H], and [La II/H]). The median of all \feh measurements, including the determination by \citet{Sousa15}, yield a value of 0.23 $\pm$ 0.08 dex, where the uncertainty represents the spread or range in abundance measurements by the different literature sources (see \citealt{Hinkel14} for more details). In terms of important planet forming materials, the median values\footnote{Individual abundance measurements can be found in the Hypatia Catalog: \url{www.hypatiacatalog.com}.} for [C/H], [O/H], [Mg/H], and [Si/H] are listed in Table \ref{tab:LitProps}, which have been normalized to the solar values of \citet{Lodders:2009p3091}. Also, the overall [$\alpha$/H] abundance for this planetary system, when utilizing the abundances from O, Mg, Si, Ca, and Ti, is 0.25 $\pm$ 0.10 dex. Converting to molar fractions, we find that the C/O ratio for this system is 0.47. 

In addition, we utilized the stellar atmospheric determinations from the previously cited spectroscopic studies to compile an estimate of the effective temperature of \thisstar. The median of the reported \teff values is 5816$\pm$90 K, where again the uncertainty represents the spread in the effective temperature measurements.  SIMBAD lists this star as a spectral type K0 dwarf based on the update of the Henry Draper Catalog \citep{Cannon:1993}, and that spectral type appears to be repeated through a number of star catalogs.  We find that the star is more appropriately considered as early G-type.  Furthermore, as the global analysis below indicates, the surface gravity of the star places it closer to the subgiant than the dwarf regime.

\subsection{Global Analysis}
\label{sec:globalfit}

We determined the parameters and uncertainties of the \thisstar planetary system using the publicly available exoplanet fitting suite, EXOFASTv2 \citep{Eastman:2013, Eastman:2019}.  We conducted a preliminary fit of the system with EXOFASTv2 to obtain an approximate measure of the stellar surface gravity, and used a loose prior on the surface gravity of the host star of $\logg = 4.0 \pm 0.25$.  We then fit the full range of available broadband photometry listed in Table \ref{tab:LitProps}, to a model spectral energy distribution (SED).  For the SED fit, we placed a Gaussian prior on the metallicity using the value of \feh\ (0.23$\pm$0.08 dex) from the available stellar spectra (see \S \ref{sec:abundances}) and the corrected Gaia parallax (see \S \ref{sec:uvw}).  We also constrain the maximum line-of-sight extinction using the dust maps of \citet{Schlegel:1998}.  The resulting SED fit provides values for $\teff = 5703$ K and R$_{\star} = 2.113$\rsun.

% We also place a Gaussian prior on R$_{\star}$ of 2.082 $\pm$ 0.019 R$_{\odot}$ from the  which was calculated following the procedures described in \citet{Stassun:2016}.  We usedwith the resulting SED fit yielding a stellar mass of $1.41\pm0.10\msun$, and a stellar radius of $2.082\pm0.019\rsun$. 

We then placed Gaussian priors on \teff, \feh, and R$_{\star}$ for the full EXOFASTv2 analysis using the values listed above.  The error limits for stellar effective temperature and thus radius are set by the accuracy of interferometric angular diameters, which show systematic differences in excess of 3\% \citep[e.g.][]{White:2018}.  We therefore adopted fractional errors of 1.5\% for \teff and 3.5\% for stellar radius, yielding priors of $\teff = 5703 \pm 86$ K and R$_{\star} = 2.113 \pm 0.74$\rsun\ for the input to the global EXOFASTv2 fit.

We used EXOFASTv2 to simultaneously model the archival ELODIE RVs, \tess photometry, and constraints on the stellar parameters from spectroscopy. Within the fit, the stellar parameters of \thisstar were determined using the MESA Isochrones and Stellar Tracks (MIST) stellar evolution models \citep{Dotter:2016, Choi:2016, Paxton:2011, Paxton:2013, Paxton:2015}. 

We fit the \tess light curve processed by the SPOC pipeline's Presearch Data Conditioning (PDC) module, which removes common-mode instrumental systematics from light curves \citep{Stumpe2012,Smith2012,Stumpe2014}. The PDC light shows some low-frequency variability (likely originating from the rotation modulations in the light curve, see Section \ref{variability}), which must be accounted for in transit modeling. It is particularly challenging to remove the low-frequency variability from the \thisstar light curve because not all of the transits have sufficient out-of-transit coverage to determine and extrapolate the variability. One of the five transits ended less than two hours before the end of Sector 15, and \tess only observed the second half of another transit after a gap in observations when the spacecraft downlinked data during its perigee passage.  We chose to model the low-frequency variability with a basis spline. 

We started by fitting a basis spline to the full \tess\ light curve, while iteratively identifying and excluding 3$\sigma$ outliers \citep{vj14}. We determined the optimal spacing between spline knots to be about 0.3 days by calculating the Bayesian Information Criterion for a series of splines fit with different knot spacings \citep{sv18}. Then, we followed a procedure similar to that of \citet{v16}, wherein we simultaneously fit the shape of the transits and the low-frequency variability (though we did not also model spacecraft systematics as is commonly done for \textit{K2} data). In brief, we performed a preliminary fit of the un-flattened \thisstar\ light curve with a \citet{mandelagol} model. Inside the fit, after each evaluation of the \citet{mandelagol} model, we fit a spline to the residuals (light curve - transit model) and subtracted it before calculating $\chi^2$, which we minimized with a Levenberg-Marquardt algorithm \citep{markwardt}. After convergence, we retrieved the best-fitting spline and subtracted it from the \tess light curve, yielding a flattened light curve for the EXOFASTv2 analysis. %We incorporated the spline variability model into the EXOFASTv2 fit in two different ways. First, w

%Though this method of flattening the light curve gives optimal results, it does not give EXOFASTv2 the ability to marginalize over different choices for removing low-frequency variability. We therefore also incorporated the spline model into the EXOFASTv2 fit directly. The implementation of this is similar to the way we incorporated the spline model into the Levenberg-Marquardt fit. At each step of the EXOFASTv2 Markov Chain Monte Carlo (MCMC) exploration, we fit a spline to the residual light curve after subtracting the transit model, and subtracted the spline before calculating the likelihood. This method requires calculating a new basis spline every MCMC step, and therefore is considerably computationally expensive, but marginalizes over any uncertainties in the flattening process. We found that both methods give consistent results, giving us confidence in our parameters.  

\begin{figure*}[!ht]
\vspace{0.3in}
\centering\includegraphics[width=0.95\linewidth, trim = 0 5.8in 0 0]{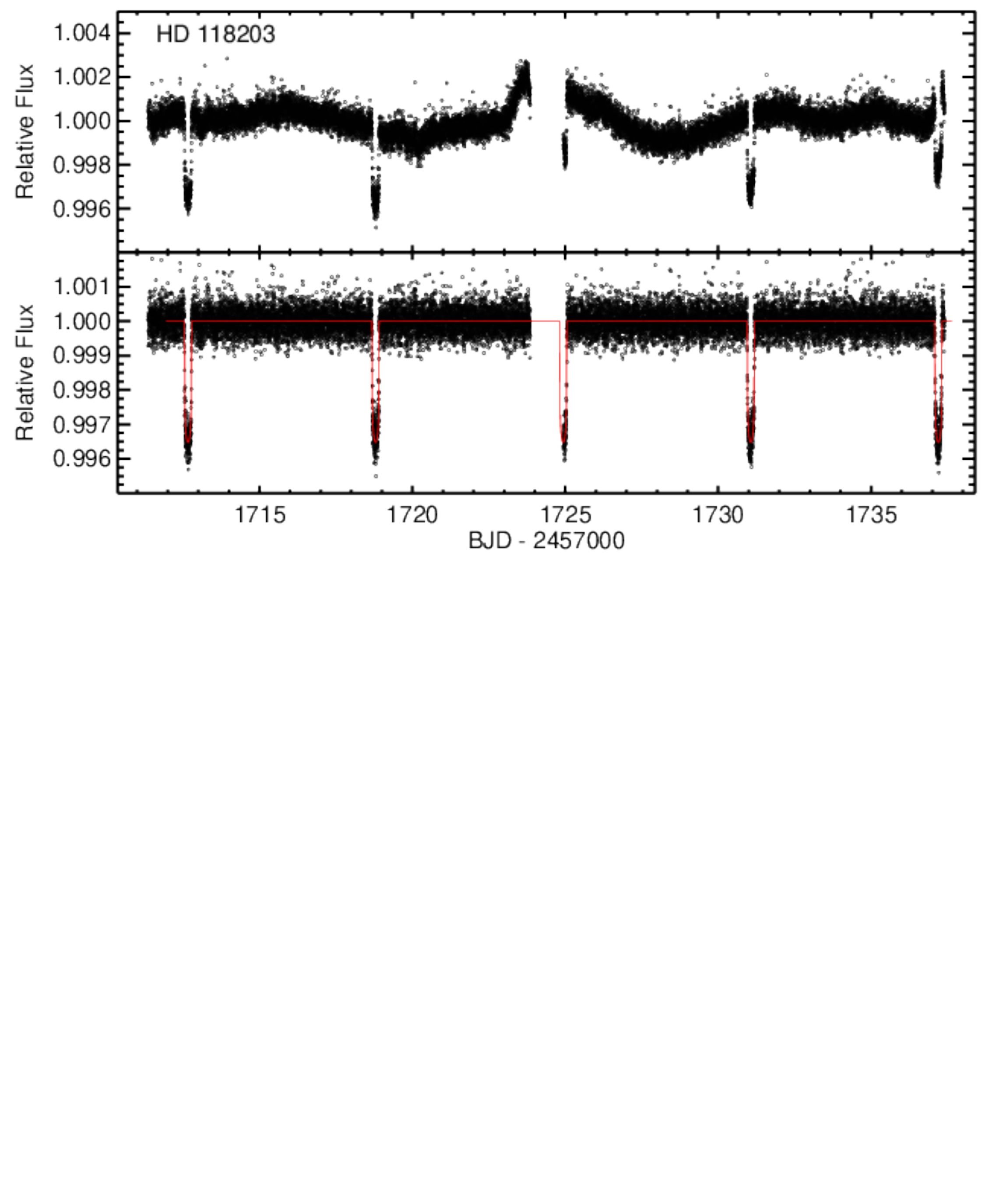}
\caption{(Top) The \tess 2-minute cadence light curve of \thisstar. (Bottom) The flattened \tess light curve used in the EXOFASTv2 fit. The observations are plotted in open black circles, and the best fit model from EXOFASTv2 is plotted in red.}
\label{figure:LC}
\vspace{9 mm}
\end{figure*}

\begin{figure}
	\centering\vspace{.0in}
	\includegraphics[width=1\linewidth, trim={0cm 23cm 0cm 0cm}, clip]{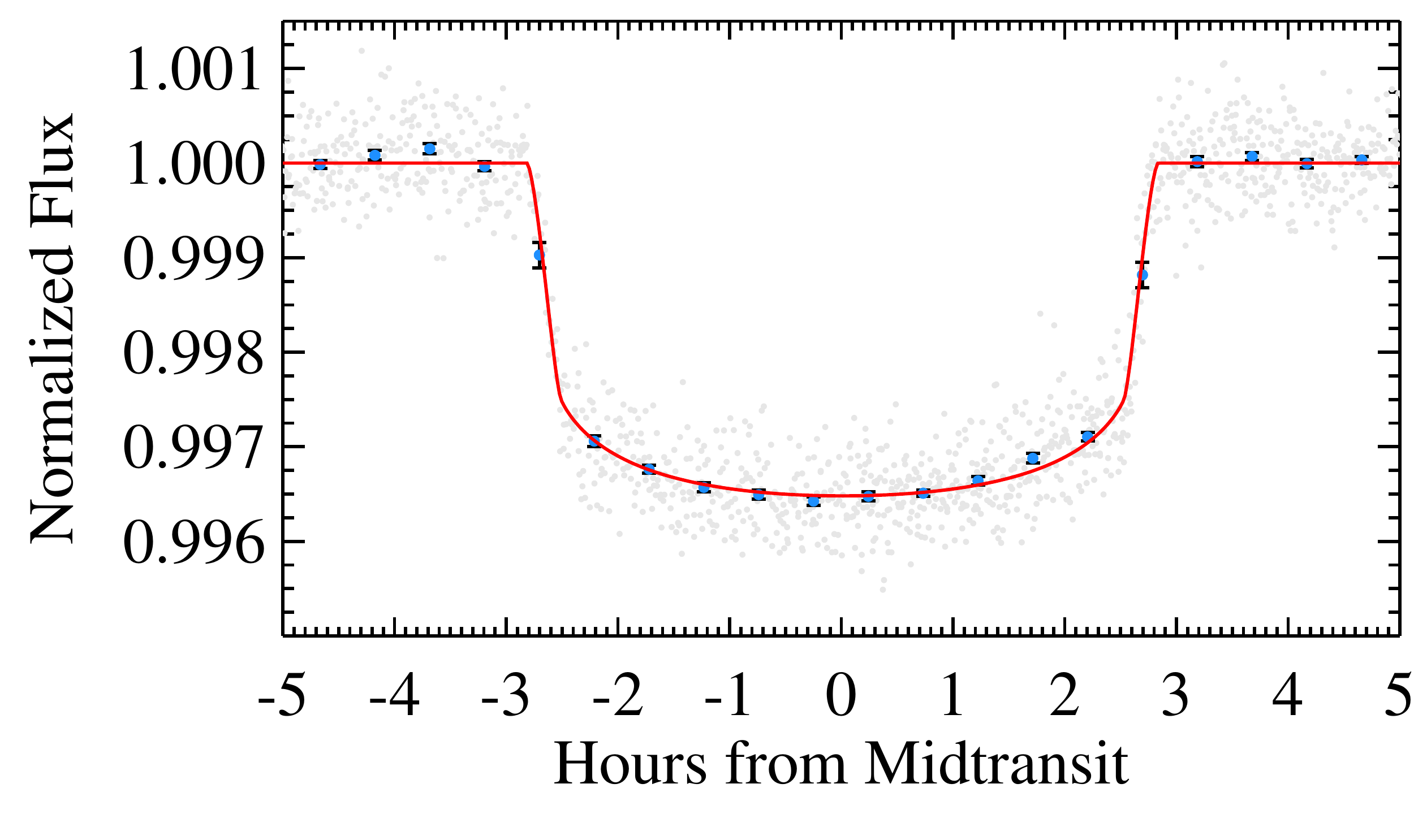}
	\caption{The \tess SPOC PDC light curve of \thisstar phase-folded to the best fit period of 6.135 days. The blue points are showing the \tess photometry in 24 minute bins. The EXOFASTv2 model is plotted in red.}
	\label{fig:zoom-in} 
\end{figure}

\begin{figure}
	\centering\vspace{.0in}
	\includegraphics[width=1\linewidth, trim={2.5cm 13cm 8.5cm 8cm}, clip]{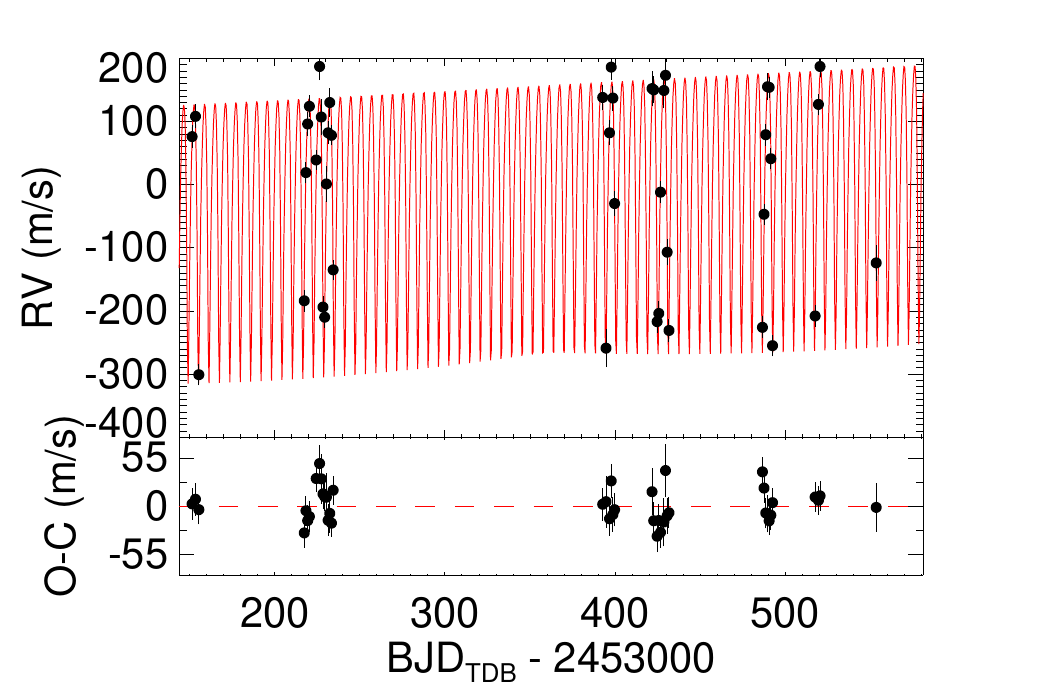}
	\includegraphics[width=1\linewidth, trim={2.5cm 13cm 8.5cm 8cm}, clip]{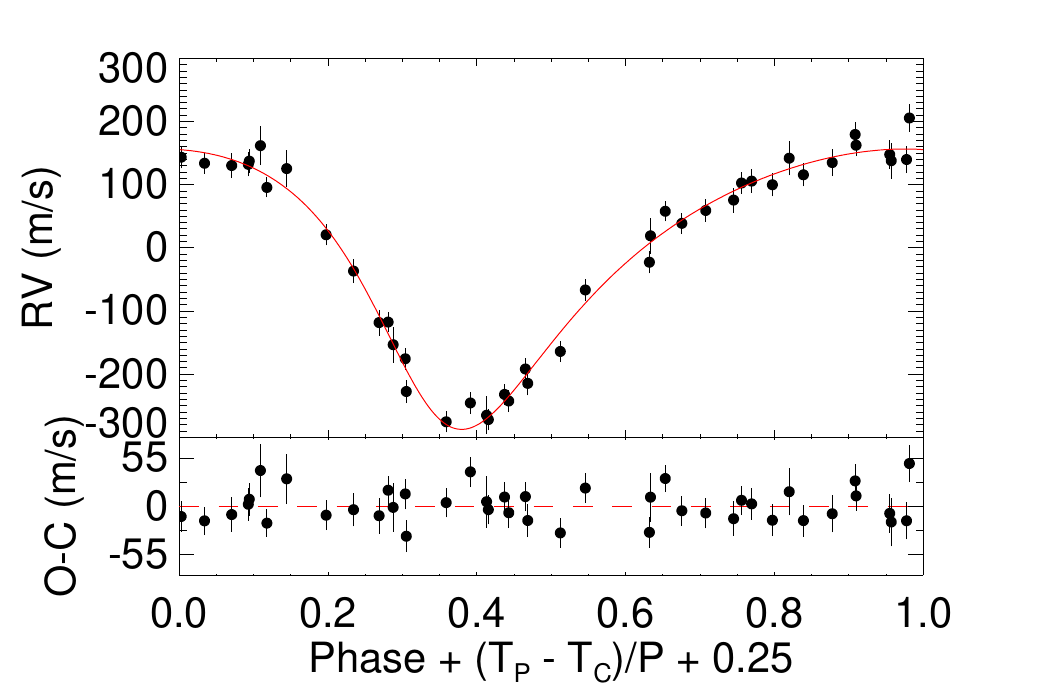}
	\caption{(Top) Radial velocity measurements from ELODIE of \thisstar \citep{daSilva2006}. (Bottom) The radial velocity measurements phase-folded to the best determined period by EXOFASTv2, 6.135 days. The EXOFASTv2 best-fit model is shown in red and the residuals of the model are shown below each plot.}
	\label{fig:RVs} 
\end{figure}

The SPOC PDC lightcurve before and after the flattening procedure is shown in Figure \ref{figure:LC}.  A phase-folded zoom-in on the transit in the flattened light curve is shown in Figure \ref{fig:zoom-in}.  We show the full and phase-folded RV curve in Figure \ref{fig:RVs}.

\begin{figure*}
	\centering\vspace{.0in}
	\includegraphics[width=1\linewidth, trim={0cm 20.5cm 0cm 7cm}, clip]{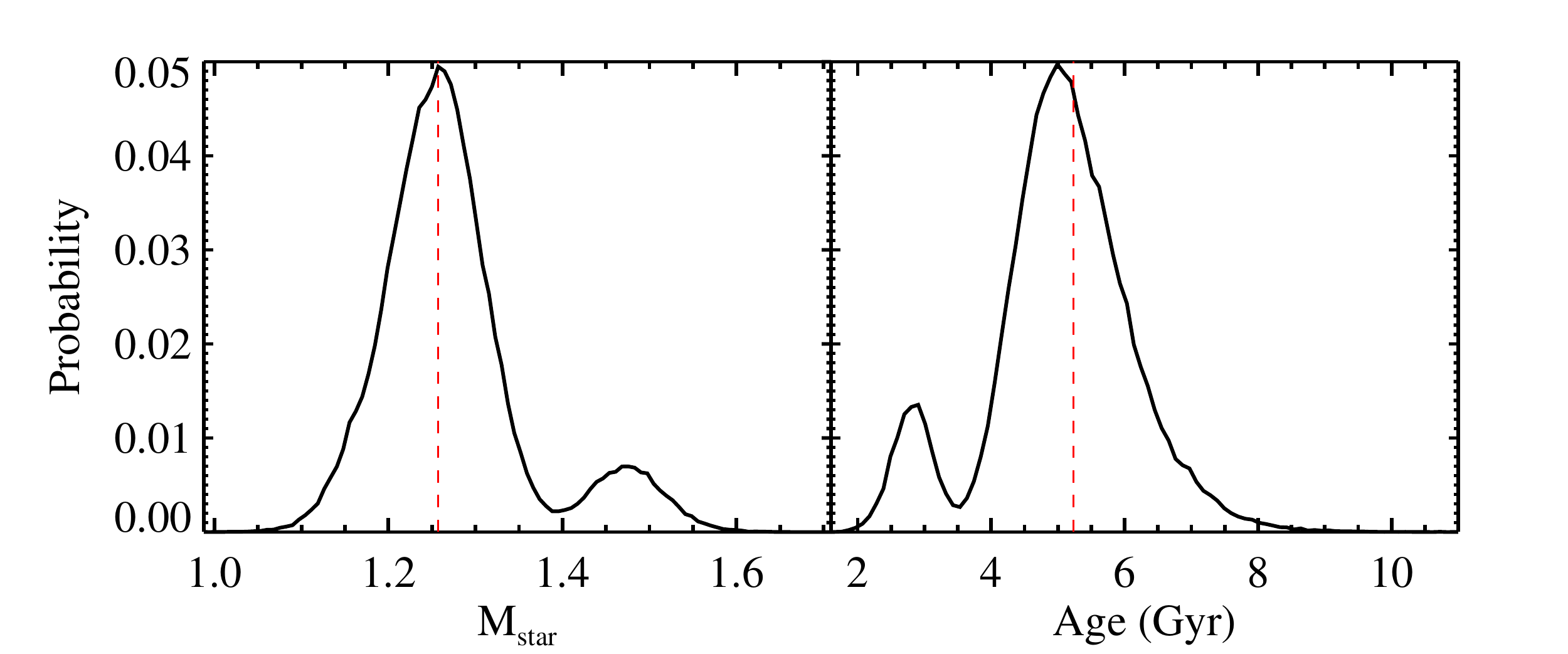}
	\caption{The probability distribution function for M$_{*}$  (Left) and Age (Right). The red line shows the median value for each parameter from the adopted solution (see \S\ref{sec:globalfit}).}
	\label{fig:PDF} 
\end{figure*}

The results of the EXOFASTv2 global fit are listed in Tables \ref{tab:exofast_stellar} and \ref{tab:exofast_planetary}. We note that the mass and age of the \thisstar\ are bimodal in our probability distribution function (PDF, see Figure \ref{fig:PDF}). The two peaks in the PDF are centered at a host star mass of 1.26 \msun\ and 1.48\msun\, corresponding to ages of 5.23 Gyr and 2.89 Gyr, respectively. In order to arrive at distinct solutions, we split the host star mass PDF at the valley between the two peaks, 1.38 \msun, and extract two separate solutions that are presented in Tables \ref{tab:exofast_stellar} and \ref{tab:exofast_planetary}. We adopt the peak at 1.26 \msun since it is significantly more probable than the 1.48 \msun solution (89.6\% compared to 10.4\%).  However both solutions are provided in Tables \ref{tab:exofast_stellar} and \ref{tab:exofast_planetary} for future work on \thisstar.  We note that the host star mass and age solutions are based on single model grid and thus do not account for systematic errors due to different input physics, which can be substantial for evolved stars (Tayar et al., in prep). Therefore, the uncertainties in mass and age reported in Table \ref{tab:exofast_stellar} are likely underestimated.

%- activity, via GALEX + any Ca II H\&K.  Possible effects on lightcurve?  Decided not to discuss this

\begin{table}
\scriptsize
\setlength{\tabcolsep}{2pt}
\centering
\caption{Literature and Measured Properties for HD~118203}
\begin{tabular}{llcc}
  \hline
  \hline
Other identifiers\dotfill & \\
\multicolumn{3}{c}{HIP 66192,   Gaia 1560420854826284928} \\
\multicolumn{3}{c}{2MASS J13340254+5343426,   TYC 3850-458-1} \\
\multicolumn{3}{c}{BD+54 1609,   TIC 286923464, TOI 1271.01}\\
\hline
\hline
Parameter & Description & Value & Source\\
\hline 
$\alpha_{J2000}$\dotfill	&Right Ascension (RA)\dotfill & 13:34:02.3894& 1	\\
$\delta_{J2000}$\dotfill	&Declination (Dec)\dotfill & +53:43:41.4752& 1	\\
\\
$l$\dotfill     & Galactic Longitude\dotfill & 109.3442934$^\circ$ & 2\\
$b$\dotfill     & Galactic Latitude\dotfill & +62.2614278$^\circ$ & 2\\
%\\
$NUV$\dotfill           & GALEX $NUV$ mag.\dotfill & 14.0481 $\pm$ 0.0059 & 3 \\
$FUV$\dotfill           & GALEX $FUV$ mag.\dotfill & 20.16 $\pm$ 0.20 & 3 \\
\\
B$_T$\dotfill			&Tycho B$_T$ mag.\dotfill & 8.903 $\pm$ 0.03 & 4	\\
V$_T$\dotfill			&Tycho V$_T$ mag.\dotfill & 8.135 $\pm$ 0.03 & 4	\\
\\
${\rm G}$\dotfill     & Gaia $G$ mag.\dotfill     & 7.8925 $\pm$ 0.05 & 1 \\
\\
J\dotfill			& 2MASS J mag.\dotfill & 6.861 $\pm$ 0.021	& 5	\\
H\dotfill			& 2MASS H mag.\dotfill & 6.608 $\pm$ 0.038	&  5	\\
K$_S$\dotfill		& 2MASS ${\rm K_S}$ mag.\dotfill & 6.543 $\pm$ 0.023 &  5	\\
\\
\textit{WISE1}\dotfill		& \textit{WISE1} mag.\dotfill & $6.472 \pm 0.078$ &  6	\\
\textit{WISE2}\dotfill		& \textit{WISE2} mag.\dotfill & $6.450 \pm 0.023$ &  6 \\
\textit{WISE3}\dotfill		& \textit{WISE3} mag.\dotfill & $6.501 \pm 0.016$ &  6	\\
\textit{WISE4}\dotfill		& \textit{WISE4} mag.\dotfill & $6.438 \pm 0.054$ &  6	\\
\\
$\mu_{\alpha}$\dotfill		& Gaia DR2 proper motion\dotfill & -85.877 $\pm$ 0.052 & 1 \\
                    & \hspace{3pt} in RA (mas yr$^{-1}$)	&  \\
$\mu_{\delta}$\dotfill		& Gaia DR2 proper motion\dotfill 	&  -78.913 $\pm$ 0.038 &  1 \\
                    & \hspace{3pt} in DEC (mas yr$^{-1}$) &  \\
$\pi$\dotfill & Gaia Parallax (mas) \dotfill & 10.810 $\pm$  0.027$^{\dagger}$ &  1 \\

$RV$\dotfill & Systemic radial velocity\dotfill  & $ -29.387\pm0.006 $  & 7 \\
%     & \hspace{3pt} velocity (\kms)  & \\
     
$d$\dotfill & Distance (pc)\dotfill & 92.26 $\pm 0.24^{\dagger}$ & 2 \\

%  UPDATE
Spec. Type\dotfill & Spectral Type\dotfill & 	 &  \\
$A_V$\dotfill & Visual extinction (mag) &  & \S\ref{sec:system}\\
%$A_V$\dotfill & Visual extinction (mag) & $ \substack{ \\ }$ & \S\ref{sec:} \\

\\
$\rm [Fe/H]$ & Iron abundance (dex) & 0.23 $\pm$ 0.08 & 8 \\
$\rm [C/H]$ & Carbon abundance (dex) & 0.31 $\pm$ 0.22 & 8 \\
$\rm [O/H]$ & Oxygen abundance (dex) & 0.30 $\pm$ 0.17 & 8 \\
$\rm [Mg/H]$ & Magnesium abundance (dex) & 0.24 $\pm$ 0.12 & 8 \\
$\rm [Si/H]$ & Silicon abundance (dex) & 0.22 $\pm$ 0.14 & 8 \\
$\rm [\alpha$/H] & $\alpha$-element (O, Mg, Si, Ca,  & 0.25 $\pm$ 0.10 & 8\\
& \hspace{3pt} and Ti) abundance (dex) &  &  \\
\\

%  UPDATE
$U^{*}$\dotfill & Space Velocity (\kms)\dotfill & $4.87 \pm 0.03$  & \S\ref{sec:uvw} \\
$V$\dotfill       & Space Velocity (\kms)\dotfill & $ -45.00 \pm 0.12$ & \S\ref{sec:uvw} \\
$W$\dotfill       & Space Velocity (\kms)\dotfill & $ 1.62 \pm 0.06$ & \S\ref{sec:uvw} \\
\hline
\end{tabular}
\begin{flushleft}
 \footnotesize{ \textbf{\textsc{NOTES:}}
 $\dagger$ Values have been corrected for the -0.82 $\mu$as offset as reported by \citet{Stassun:2019}.\\
 $*$ $U$ is in the direction of the Galactic center. \\
 References are: $^1$\citet{Gaia:2018}, $^2$\citet{Stassun:2019}, $^3$\citet{GALEX:2017},  $^4$\citet{Hog:2000}, $^5$\citet{Cutri:2003}, $^6$\citet{Cutri:2014}, $^7$\citet{daSilva2006}, $^8$Hypatia Catalog \citep{Hinkel14}, \url{www.hypatiacatalog.com}
}
\end{flushleft}
\label{tab:LitProps}
\end{table}

\label{tab:star} %this is cited in a few places within the manuscript

\begin{table*}
\scriptsize
\centering
\caption{Median values and 68\% confidence intervals for the global model of \thisstar}
\begin{tabular}{llcccccc}
  \hline
  \hline
Parameter & Units & {\bf Values (Adopted Solution)} & Values (Secondary Solution) & & \\
\hline
\multicolumn{2}{l}{Stellar Parameters:}&\smallskip\\
Probability\dotfill & &89.6\% & 10.4\% \\
~~~~$M_*$\dotfill &Mass (\msun)\dotfill &$1.257^{+0.051}_{-0.056}$&$1.477^{+0.045}_{-0.043}$\\
~~~~$R_*$\dotfill &Radius (\rsun)\dotfill &$2.103^{+0.055}_{-0.054}$&$2.168^{+0.053}_{-0.051}$\\
~~~~$L_*$\dotfill &Luminosity (\lsun)\dotfill &$4.18^{+0.35}_{-0.33}$&$4.66^{+0.36}_{-0.34}$\\
~~~~$\rho_*$\dotfill &Density (g cm$^{-3}$)\dotfill &$0.190\pm0.012$&$0.205\pm0.012$\\
~~~~$\log{g}$\dotfill &Surface gravity (g cm$^{-2}$)\dotfill &$3.891^{+0.019}_{-0.020}$&$3.936^{+0.016}_{-0.017}$\\
~~~~$T_{\rm eff}$\dotfill &Effective Temperature (K)\dotfill &$5692\pm83$&$5761^{+77}_{-78}$\\
~~~~$[{\rm Fe/H}]$\dotfill &Metallicity (dex)\dotfill &$0.223\pm0.073$&$0.265^{+0.070}_{-0.068}$\\
~~~~$[{\rm Fe/H}]_{0}^\dagger$\dotfill &Initial Metallicity \dotfill &$0.225^{+0.069}_{-0.071}$&$0.285^{+0.062}_{-0.058}$\\
~~~~$Age$\dotfill &Age (Gyr)\dotfill &$5.23^{+0.90}_{-0.70}$&$2.89^{+0.33}_{-0.32}$\\
~~~~$EEP^\ddagger$\dotfill &Equal Evolutionary Phase \dotfill &$457.0^{+3.9}_{-3.7}$&$407.7^{+5.4}_{-6.8}$\\
~~~~$\dot{\gamma}$\dotfill &RV slope (m/s/day)\dotfill &$0.139^{+0.026}_{-0.027}$&$0.138^{+0.026}_{-0.027}$\\
\hline
\end{tabular}
\begin{flushleft} 
  \footnotesize{ 
    \textbf{\textsc{NOTES:}}
$^\dagger$The initial metallicity is the metallicity of the star when it was formed.\\
$^\ddagger$The Equal Evolutionary Point corresponds to static points in a stars evolutionary history when using the MIST isochrones and can be a proxy for age. See \S2 in \citet{Dotter:2016} for a more detailed description of EEP.
               }
 \end{flushleft}
\label{tab:exofast_stellar}
\end{table*}
\begin{table*}
\scriptsize
\centering
\caption{Median values and 68\% confidence intervals for the global model of \thisstar}
\begin{tabular}{llcccc}
  \hline
  \hline
Parameter & Description (Units) & {\bf Values (Adopted Solution)} & Values (Secondary Solution) & & \\
\hline
Probability\dotfill & &89.6\% & 10.4\% \\
~~~~$P$\dotfill &Period (days)\dotfill &$6.134980^{+0.000038}_{-0.000037}$&$6.134993^{+0.000038}_{-0.000037}$\\
~~~~$R_P$\dotfill &Radius (\rj)\dotfill &$1.133^{+0.031}_{-0.030}$&$1.167^{+0.029}_{-0.028}$\\
~~~~$M_P$\dotfill &Mass (\mj)\dotfill &$2.173^{+0.077}_{-0.080}$&$2.417^{+0.073}_{-0.068}$\\
~~~~$T_C$\dotfill &Time of conjunction (\bjdtdb)\dotfill &$2458712.66156^{+0.00023}_{-0.00025}$&$2458712.66155^{+0.00022}_{-0.00023}$\\
~~~~$T_0^\dagger$\dotfill &Optimal conjunction Time (\bjdtdb)\dotfill &$2458724.93152^{+0.00022}_{-0.00024}$&$2458724.93154\pm0.00021$\\
~~~~$a$\dotfill &Semi-major axis (AU)\dotfill &$0.07082^{+0.00095}_{-0.0011}$&$0.07474\pm0.00074$\\
~~~~$i$\dotfill &Inclination (Degrees)\dotfill &$88.75^{+0.86}_{-1.0}$&$89.12^{+0.62}_{-0.85}$\\
~~~~$e$\dotfill &Eccentricity \dotfill &$0.316\pm0.021$&$0.303\pm0.021$\\
~~~~$\omega_*$\dotfill &Argument of Periastron (Degrees)\dotfill &$153.6^{+3.5}_{-3.6}$&$156.1^{+3.6}_{-3.7}$\\
~~~~$T_{eq}$\dotfill &Equilibrium temperature (K)\dotfill &$1496\pm26$&$1496^{+26}_{-25}$\\
~~~~$\tau_{\rm circ}$\dotfill &Tidal circularization timescale (Gyr)\dotfill &$12.7^{+1.5}_{-1.4}$&$13.6^{+1.5}_{-1.4}$\\
~~~~$K$\dotfill &RV semi-amplitude (m/s)\dotfill &$218.3^{+5.2}_{-5.1}$&$217.0\pm5.0$\\
%~~~~$\log{K}$\dotfill &Log of RV semi-amplitude \dotfill %&$2.339\pm0.010$&$2.3364^{+0.0099}_{-0.010}$\\
~~~~$R_P/R_*$\dotfill &Radius of planet in stellar radii \dotfill &$0.05538^{+0.00023}_{-0.00022}$&$0.05534^{+0.00021}_{-0.00020}$\\
~~~~$a/R_*$\dotfill &Semi-major axis in stellar radii \dotfill &$7.24\pm0.15$&$7.42\pm0.15$\\
~~~~$\delta$\dotfill &Transit depth (fraction)\dotfill &$0.003067^{+0.000026}_{-0.000024}$&$0.003063^{+0.000023}_{-0.000022}$\\
%~~~~$Depth$\dotfill &Flux decrement at mid transit \dotfill %&$0.003067^{+0.000026}_{-0.000024}$&$0.003063^{+0.000023}_{-0.000022}$\\
~~~~$\tau$\dotfill &Ingress/egress transit duration (days)\dotfill &$0.01259^{+0.00052}_{-0.00019}$&$0.01249^{+0.00034}_{-0.00011}$\\
~~~~$T_{14}$\dotfill &Total transit duration (days)\dotfill &$0.23516^{+0.00069}_{-0.00062}$&$0.23508^{+0.00061}_{-0.00058}$\\
~~~~$T_{FWHM}$\dotfill &FWHM transit duration (days)\dotfill &$0.22246^{+0.00054}_{-0.00052}$&$0.22249\pm0.00053$\\
~~~~$b$\dotfill &Transit Impact parameter \dotfill &$0.125^{+0.10}_{-0.086}$&$0.092^{+0.089}_{-0.065}$\\
~~~~$b_S$\dotfill &Eclipse impact parameter \dotfill &$0.17^{+0.13}_{-0.11}$&$0.119^{+0.11}_{-0.084}$\\
~~~~$\tau_S$\dotfill &Ingress/egress eclipse duration (days)\dotfill &$0.01698^{+0.00091}_{-0.00075}$&$0.01617^{+0.00073}_{-0.00065}$\\
~~~~$T_{S,14}$\dotfill &Total eclipse duration (days)\dotfill &$0.309\pm0.012$&$0.299^{+0.012}_{-0.011}$\\
~~~~$T_{S,FWHM}$\dotfill &FWHM eclipse duration (days)\dotfill &$0.292\pm0.012$&$0.283\pm0.011$\\
~~~~$\delta_{S,3.6\mu m}$\dotfill &Blackbody eclipse depth at 3.6$\mu$m (ppm)\dotfill &$227.2^{+8.9}_{-8.5}$&$223.0^{+8.5}_{-8.1}$\\
~~~~$\delta_{S,4.5\mu m}$\dotfill &Blackbody eclipse depth at 4.5$\mu$m (ppm)\dotfill &$308.7^{+10.}_{-9.4}$&$303.3^{+9.5}_{-9.0}$\\
~~~~$\rho_P$\dotfill &Density (g cm$^{-3}$)\dotfill &$1.85\pm0.13$&$1.89^{+0.13}_{-0.12}$\\
~~~~$logg_P$\dotfill &Surface gravity \dotfill &$3.622\pm0.021$&$3.644\pm0.020$\\
~~~~$\Theta$\dotfill &Safronov Number \dotfill &$0.2160^{+0.0074}_{-0.0072}$&$0.2093^{+0.0067}_{-0.0065}$\\
~~~~$\fave$\dotfill &Incident Flux (\fluxcgs)\dotfill &$1.031^{+0.071}_{-0.067}$&$1.039^{+0.070}_{-0.066}$\\
~~~~$T_P$\dotfill &Time of Periastron (\bjdtdb)\dotfill &$2458707.116^{+0.048}_{-0.045}$&$2458707.163^{+0.050}_{-0.048}$\\
~~~~$T_S$\dotfill &Time of eclipse (\bjdtdb)\dotfill &$2458708.495\pm0.082$&$2458708.521\pm0.082$\\
~~~~$T_A$\dotfill &Time of Ascending Node (\bjdtdb)\dotfill &$2458710.981^{+0.060}_{-0.062}$&$2458710.942^{+0.059}_{-0.062}$\\
~~~~$T_D$\dotfill &Time of Descending Node (\bjdtdb)\dotfill &$2458707.344^{+0.041}_{-0.040}$&$2458707.376^{+0.041}_{-0.040}$\\
~~~~$e\cos{\omega_*}$\dotfill & \dotfill &$-0.282\pm0.022$&$-0.276\pm0.022$\\
~~~~$e\sin{\omega_*}$\dotfill & \dotfill &$0.140\pm0.019$&$0.123^{+0.019}_{-0.018}$\\
~~~~$M_P\sin i$\dotfill &Minimum mass (\mj)\dotfill &$2.173^{+0.077}_{-0.080}$&$2.417^{+0.073}_{-0.068}$\\
~~~~$M_P/M_*$\dotfill &Mass ratio \dotfill &$0.001653^{+0.000044}_{-0.000042}$&$0.001563^{+0.000036}_{-0.000037}$\\
~~~~$d/R_*$\dotfill &Separation at mid transit \dotfill &$5.71^{+0.26}_{-0.25}$&$6.00\pm0.26$\\
~~~~$P_T$\dotfill &A priori non-grazing transit prob \dotfill &$0.1654^{+0.0076}_{-0.0071}$&$0.1574^{+0.0071}_{-0.0065}$\\
~~~~$P_{T,G}$\dotfill &A priori transit prob \dotfill &$0.1848^{+0.0085}_{-0.0079}$&$0.1758^{+0.0079}_{-0.0072}$\\
~~~~$P_S$\dotfill &A priori non-grazing eclipse prob \dotfill &$0.1246^{+0.0031}_{-0.0027}$&$0.1230^{+0.0028}_{-0.0024}$\\
~~~~$P_{S,G}$\dotfill &A priori eclipse prob \dotfill &$0.1392^{+0.0035}_{-0.0030}$&$0.1374^{+0.0031}_{-0.0027}$\\
\smallskip\\\multicolumn{2}{l}{Wavelength Parameters:}&TESS\smallskip\\
~~~~$u_{1}$\dotfill &linear limb-darkening coeff \dotfill &$0.294\pm0.028$&$0.290\pm0.028$\\
~~~~$u_{2}$\dotfill &quadratic limb-darkening coeff \dotfill &$0.206\pm0.043$&$0.212^{+0.044}_{-0.043}$\\
\smallskip\\\multicolumn{2}{l}{Telescope Parameters:}&ELODIE\smallskip\\
~~~~$\gamma_{\rm rel}$\dotfill &Relative RV Offset (m/s)\dotfill &$-29339.4^{+3.4}_{-3.3}$&$-29339.5\pm3.3$\\
~~~~$\sigma_J$\dotfill &RV Jitter (m/s)\dotfill &$16.3^{+3.5}_{-3.0}$&$16.2^{+3.5}_{-3.1}$\\
~~~~$\sigma_J^2$\dotfill &RV Jitter Variance \dotfill &$267^{+130}_{-90}$&$263^{+130}_{-90}$\\
\smallskip\\\multicolumn{2}{l}{Transit Parameters:}&TESS UT 2019-S1-5. (TESS)\smallskip\\
~~~~$\sigma^{2}$\dotfill &Added Variance \dotfill &$0.0000000169^{+0.0000000036}_{-0.0000000035}$&$0.0000000170\pm0.0000000036$\\
~~~~$F_0$\dotfill &Baseline flux \dotfill &$1.0000022\pm0.0000080$&$1.0000021\pm0.0000079$\\
\hline
\end{tabular}
\begin{flushleft} 
  \footnotesize{ 
    \textbf{\textsc{\hspace{0.75in}NOTES:}}
See Table 3 in \citet{Eastman:2019} for a list of the derived and fitted parameters in EXOFASTv2.\\
$^\dagger$Minimum covariance with period.
All values in this table for the secondary eclipse of \thisplanet are predicted values from our global analysis.               
               }
 \end{flushleft}
\label{tab:exofast_planetary}
\end{table*}

As an additional check on the system parameters, we employed the Bayesian code PARAM \citep{PARAM,PARAM2,PARAM3} to determine fundamental properties of \thisstar following a grid-based approach, whereby observed quantities (namely, $T_{\rm eff}$, \feh, $\pi$, and apparent magnitudes) were matched to a well-sampled grid of stellar evolutionary tracks. The optimization method we adopted, the so-called 1-step approach, takes into account the entire set of input parameters at once in order to compute the PDFs for the stellar properties. This method is an updated version of the \citet{PARAM2} implementation, in which the code considers both the apparent and model-derived absolute magnitudes, as part of an additional step, to derive the extinction and distance to the star (2-step approach). The underlying grid of stellar evolutionary tracks (and relevant physical inputs) on which we ran PARAM is described in Section~2 of \citet{PARAM3}. Element diffusion has, however, now been included (leading to a different He-enrichment ratio of $\Delta Y/\Delta Z = 1.33$, where $Y$ and $Z$ are respectively the initial helium and metal mass fractions). We note that the PARAM output is consistent with the adopted EXOFASTv2 solution.

In addition to using EXOFASTv2 to model the system, we performed an additional analysis of the \tess light curve and ELODIE RV data using \texttt{allesfitter} \citep[][and in prep.]{GuentherDaylan2019}. \texttt{allesfitter} is an analysis framework that allows the orbital and dynamical parameters of a multi-body system to be inferred given some RV and photometric data. In this work, we omit a complete discussion of \texttt{allesfitter} and we refer the reader to \citet{Guenther2019a, Guenther2019b,Daylan2019} for its implementation details. Using a Gaussian process to model the background as part of this analysis, we find a radius ratio of $5.53\% \pm 0.03$ and an eccentricity of $0.30 \pm 0.03$. Although we select the EXOFASTv2 results as the final system solution, the fact that the system parameters found by \texttt{allesfitter} are in agreement is additional verification of the robustness of the global fit.

\vspace{16 mm}
\subsection{Stellar Variability}\label{variability}

To search for signs of stellar variability, we visually inspected the SPOC Simple Aperture Photometry (SAP) version \citep{Twicken:2010,Morris:2017} of the \tess lightcurve from Sector~15 and found a potentially periodic signal with an amplitude of around 0.6\% and an estimated period of 20--25 days.
%\textcolor{red}{(should we include a figure with SAP lightcurve?)}. [From Josh: probably not needed for now, but we can add it in later.]
The spectroscopically measured projected rotational velocity of the host star of 4.7\kms \citep{daSilva2006}, coupled with the newly obtained stellar radius of 2.1~\rsun\ (see Table \ref{tab:exofast_stellar}), implies a stellar rotational period of 22 days under the assumption that the stellar rotation axis is orthogonal to the line of sight. This suggests that the observed variability could be caused by the stellar rotation and the presence of starspots. With the pending release of the \tess light curves from Sector 16, it should be possible to confirm the astrophysical origin as well as the periodic nature of this signal.

\thisstar was also observed by the KELT-North Telescope \citep{Pepper2007} between 2012 February 19 and 2014 December 30, which provided a total of 4221 photometric data points. We performed a Lomb--Scargle analysis on KELT photometry and did not detect any periodic astrophysical signals with a semi-amplitude upper limit of 24\thinspace ppt, equal to the standard deviation of the raw KELT lightcurve. Given that the apparent variability amplitude from the \tess photometry is smaller than the detection threshold for KELT, the non-detection of both the rotation signal and the transit in KELT is unsurprising.

\subsection{Location and Kinematics in the Galaxy}
\label{sec:uvw}

\thisstar is labeled as a high-proper motion star by SIMBAD.  This could be due to the fact that it is relatively close, or it could be because it has higher-than average space velocity for a typical thin disk star (which might imply that it is older than the typical age of the thin disk or a member of the thick disk), or some combination of the two. To determine the location and kinematics of \thisstar, we use the Gaia parallax (corrected for the 82 $\mu$as systematic according to \citealt{Stassun:2018}), and the Gaia proper motions and parallax \citep{Gaia:2018}.   

The corrected parallax is $10.90201 \pm 0.0275~\rm mas$, which implies a distance from the sun of of $91.7\pm0.2$~pc.  \footnote{Given the very small (0.3\%) fractional uncertainty in the parallax, we do not attempt to correct for Lutz-Kelker bias \citep{Lutz:1973}.} The Galactic coordinates of \thisstar are $(\ell,b)=(109.34^\circ,+62.3^\circ)$, and thus the difference in the vertical position of \thisstar from that of the sun is $Z-Z_\odot = 81$~pc.  Given the $Z_\odot \simeq 30$~pc determined from giants in the local solar neighborhood \citep{bovy:2017}, we find $Z \simeq 120~$pc.  This is comparable to the scale height of early G stars in the local Galactic disk as determined by \citet{bovy:2017}.  Given the relatively small distance of \thisstar from the Sun and the fact that it is located at roughly $\ell \sim 90^\circ$, we find that the Galactocentric distance of \thisstar is essentially the same as that of the Sun (to within $\la 1\%$).  

We determine the space velocity of \thisstar to be $(U,V,W)=(4.87\pm 0.03, -45.00\pm 0.12, 1.62 \pm 0.06)~{\rm km~s^{-1}}$, correcting for the velocity of the Sun with respect to the local standard of rest as determined by \citet{Coskunoglu:2011}. Thus \thisstar has relatively small $U$ and $W$ velocities relative the dispersion in the local disk, but a relatively high asymmetric drift.  This generally indicates a relatively old (but still thin disk) star, which would not be surprising given that it is a mid-type star.  The classification scheme of \citet{Bensby:2003} gives a $97.5\%$ chance that this is a thin disk star.  This is corroborated by the abundances, $\feh = 0.23 \pm 0.08$ and [$\alpha$/Fe] $= 0.25 \pm 0.10$. 

%%%%%%%%%%%%%%%%%%%%%%%%%%%%%%%%%%%%%%%%%%%%%%%%%%%%%

\section{Discussion}
\label{sec:disc}

%%%%%%%%%%%%%%%%%%%%%%%%%%%%%%%%%%%%%%%%%%%%%%%%%%%%%

\subsection{Demographic Properties}

The fact that \thisplanet orbits such a bright host star makes this an exciting representative of a transiting planet.  Out of the 3074 confirmed planets listed on the NASA Exoplanet Archive\footnote{https://exoplanetarchive.ipac.caltech.edu, queried on 14 October 2019} that are not labeled as ``controversial", that are known to transit, and that have a host star with a recorded optical magnitude, \thisstar is brighter than all but 12 host stars, and is the 8th brightest transit host in the northern hemisphere.

\begin{figure*}
	\centering\vspace{.0in}
	\includegraphics[width=0.95\linewidth, trim = 0 0 0 0]{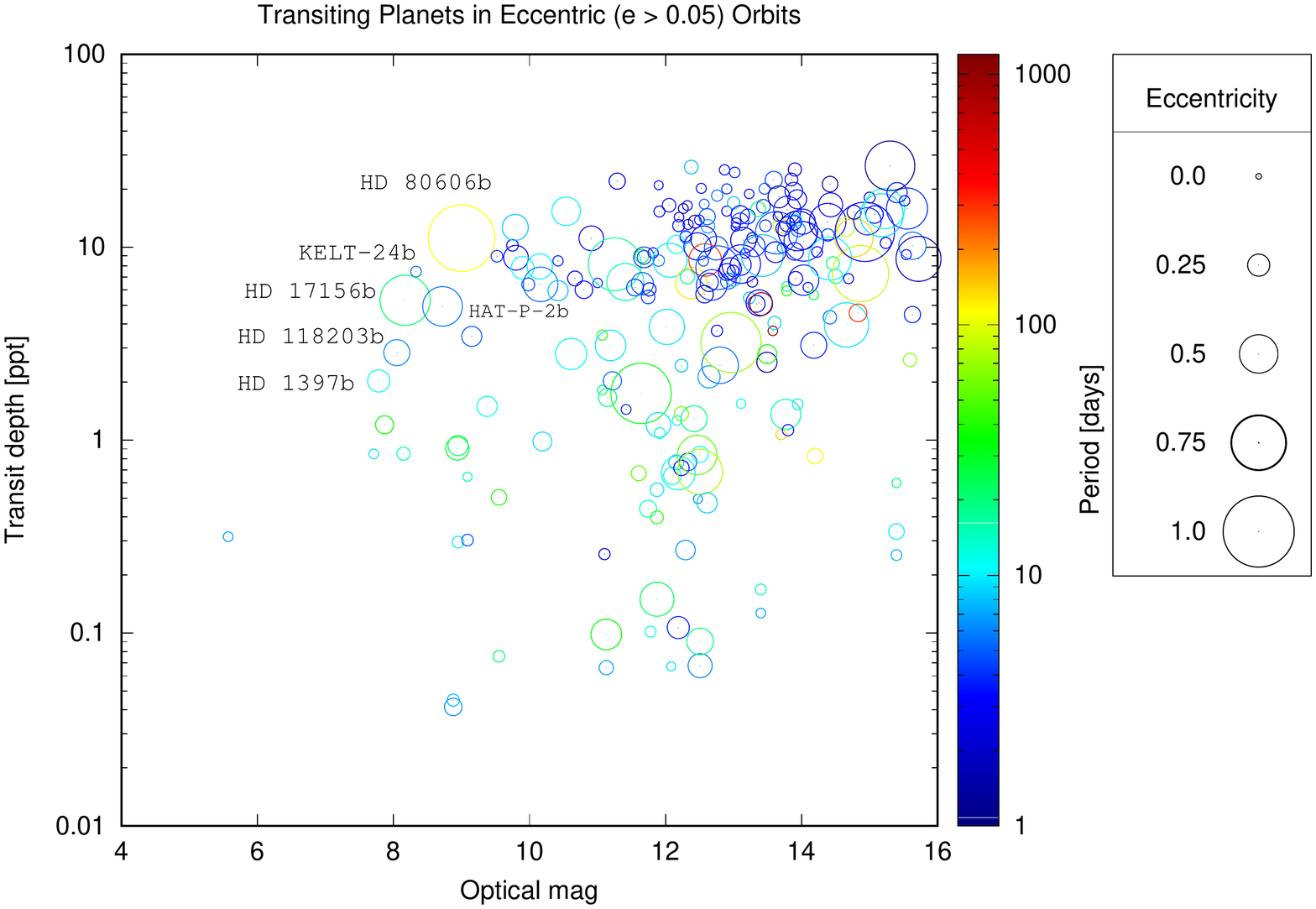}
	\caption{All known transiting planets in orbits with eccentricity greater than 0.05.  \thisplanet occupies a spot in the upper left of the distribution, along with a few other transiting planets orbiting bright stars with large transit depths.  Data from the NASA Exoplanet Archive, retrieved 15 October 2019.}
	\label{fig:pop_fig} 
\end{figure*}

\begin{figure*}
	\centering\vspace{.0in}
	\includegraphics[width=0.95\linewidth, trim = 0 0 0 0]{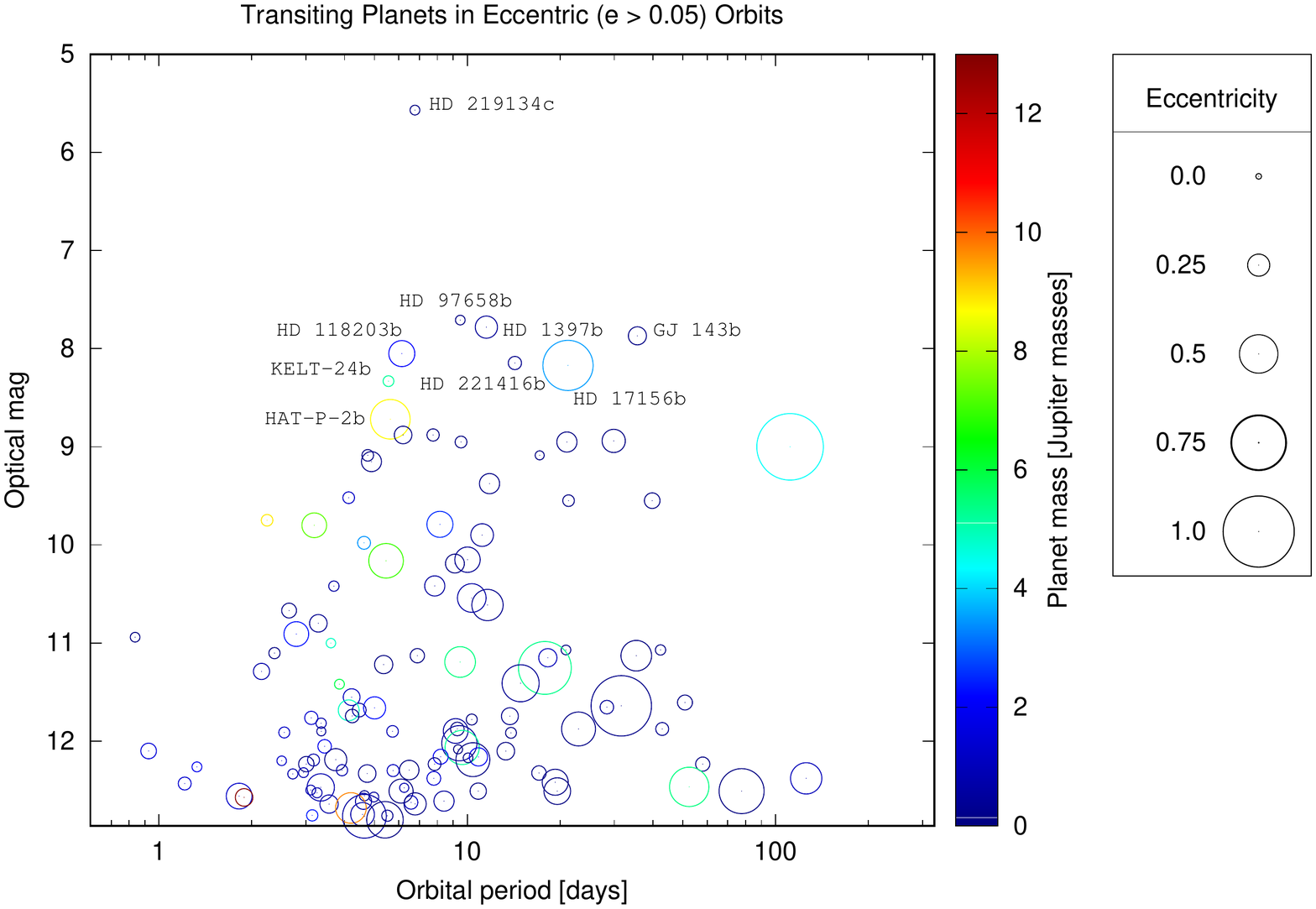}
	\caption{All known transiting planets in orbits with eccentricity greater than 0.05.  \thisplanet occupies a spot near the top of the distribution, along with a few other transiting planets orbiting bright stars.  Data from the NASA Exoplanet Archive, retrieved 15 October 2019.}
	\label{fig:pop_fig_2} 
\end{figure*}

Additionally, \thisplanet is a massive transiting planet in a short-period, eccentric orbit.  That combination of properties is shown in Figures \ref{fig:pop_fig} and \ref{fig:pop_fig_2}, which display all known transiting planets in orbits with significant eccentricity ($e>0.05$).  \thisplanet is one of the few transiting planets in eccentric orbits with a bright host star.

%%%%%%%%%%%%%%%%%%%%%%%%%%%%%%%%%%%%%%%%%%%%%%%%%%%%%

\subsection{Expectations for Transits of RV-detected Planets}

By observing bright stars covering almost the entire sky, \tess offers a unique opportunity to search for transits of exoplanets that were discovered via RV variations of their host stars. Considering the geometric transit probability of each RV-detected system and the observational strategy of \tess, \citet{Dalba2019} predicted that \tess would observe transits for $\sim$11 RV-detected planets in its primary mission. However, only three of these detections were expected to be novel, such that the RV-detected planet was not previously known to transit. 
%Since that prediction was made, the observational strategy of \tess has been altered to reduce photometric contamination due to scattered light. As a result, certain stars (e.g., \thisstar) receive more observation than originally expected while other stars are not observed at all. Although a new yield prediction given the updated observational strategy has not been made, the uniform distribution of RV-detected planets on the sky suggests that the yield of transits is unlikely to change. 
 
The discovery of transits for \thisplanet at approximately halfway through the \tess primary mission is consistent with this prediction. If the detection rate is (roughly) one per cycle, then we should expect one more discovery by the end of Cycle 2.  As is the case with \thisplanet, any other RV-detected planets found to transit are likely to have short-period orbits compared to the average RV-detected planet.  

Since the orbital period of \thisplanet is shorter than the duration of a \tess sector, its a priori transit probability is not reduced due to the observational baseline.  The a priori geometric transit probability of \thisplanet is 0.21$\pm$0.02 \citep{Dalba2019}, placing it in the 98th percentile among all RV-detected exoplanets.  The a posteriori transit probability \citep{Stevens:2013} may be higher; however this depends on the true mass distribution of planets with masses within a factor of a few of \thisplanet.  Based on the fact that, from RV surveys, it is known that companions with minimum masses somewhat larger than that of Jupiter on relatively short period ($P \la$ a few years) orbits appear to have a mass function that decreases with increasing mass, at least until the `driest' part of the brown dwarf desert \citep{grether:2006}, the a posteriori transit probability is likely to be higher for planets in this minimum mass regime than the naive a priori transit probability \citep{Stevens:2013} would suggest.  Indeed, the population synthesis models used by \citet{Stevens:2013} to determine the scale factor that relates the a posteriori and a priori transit probabilities suggest that \thisplanet is right on the boundary of having a significant boost in the probability of transit relative to the naive a priori estimate.  In general, we follow \citet{Stevens:2013} and suggest that targeting RV-detected planets with minimum mass in the regime where the true mass function is likely decreasing sharply with increasing mass (super Jupiters, super Earths and Neptunes) may result in a higher yield of transiting planets than naive a priori transit estimates would imply.  There are more than 30 RV-detected exoplanets with a priori transit probabilities as observed by \tess between 0.1 and 0.3 \citep[][their Table 1]{Dalba2019}. Although some of these are already known to transit (e.g., 55~Cnc~e), the next RV-detected exoplanet found to transit likely resides in this group, and is even more likely to be in the minimum mass regimes mentioned above.  \citet{Stevens:2013} also provide a list of particularly promising systems with transit probabilities that are likely to be higher than naively expected (see their Table 3). 

%%%%%%%%%%%%%%%%%%%%%%%%%%%%%%%%%%%%%%%%%%%%%%%%%%%%%

\subsection{Observing Prospects}

\thisstar was observed by \tess in Sector 15 (2019 Aug 15 to 2019 Sep 11) which is the data set we have analyzed here.  The star was also observed in \tess Sector 16 (2019 Sep 11 to 2019 Oct 07), although that data is not yet available at the time of this writing.  It is also expected to be observed in \tess Sector 22 (2020 Feb 18 to 2020 Mar 18).  While we do not expect those data to lead to significant updates to the fundamental physical parameters of the system, the additional photometry can potentially provide a more precise measurement of the rotational period of the host star.

Another exciting prospect is that the additional photometry could allow the detection of solar-like oscillations in \thisstar using asteroseismic analysis.  That is primarily due to that star's brightness of $T=7.45$ and the fact that it is slightly evolved (see \S \ref{sec:globalfit}). \thisstar is therefore within the regime suitable for asteroseismology described by \citet{Campante16}. The actual observability of solar-like oscillations awaits a more detailed analysis of the \tess photometry once the future sectors of observations are acquired.

The combination of a relatively short orbital period, bright host star, and eccentric orbit presents an opportunity for phase curve observations of the system.  The secondary eclipse is predicted to take place 33 hours after periastron passage.  Infrared observations of the system during and after periastron passage through the secondary eclipse could provide insight into the thermal properties of the planetary atmosphere and dynamics of heat transport, such as observed for HAT-P-2b \citep{Lewis:2013}.  Although {\it Spitzer} observations are no longer available, JWST presents an excellent opportunity for such observations.

%%%%%%%%%%%%%%%%%%%%%%%%%%%%%%%%%%%%%%%%%%%%%%%%%%%%%

\section*{Acknowledgements}

We would like to thank Avi Shporer and Chelsea Huang for helpful conversations. 

This paper includes data collected with the \tess mission, obtained from the MAST data archive at the Space Telescope Science Institute (STScI). Funding for the \tess mission is provided by the NASA Explorer Program. STScI is operated by the Association of Universities for Research in Astronomy, Inc., under NASA contract NAS 5–26555.

TD acknowledges support from MIT's Kavli Institute as a Kavli postdoctoral fellow. AV's work was performed under contract with the California Institute of Technology / Jet Propulsion Laboratory funded by NASA through the Sagan Fellowship Program executed by the NASA Exoplanet Science Institute.

DH acknowledges support by the National Aeronautics and Space Administration (80NSSC18K1585, 80NSSC19K0379) awarded through the TESS Guest Investigator Program and by the National Science Foundation (AST-1717000).

TLC acknowledges support from the European Union's Horizon 2020 research and innovation programme under the Marie Sk\l{}odowska-Curie grant agreement No.~792848 (PULSATION). This work was supported by FCT/MCTES through national funds (UID/FIS/04434/2019).

DB is supported in the form of work contract FCT/MCTES through national funds and by FEDER through COMPETE2020 in connection to these grants: UID/FIS/04434/2019; PTDC/FIS-AST/30389/2017 \& POCI-01-0145-FEDER-030389.

JAB acknowledges support from NASA's Exoplanet Exploration Program Office.

This work has made use of NASA's Astrophysics Data System, the SIMBAD database operated at CDS, Strasbourg, France, and the VizieR catalogue access tool, CDS, Strasbourg, France.  
We acknowledge the use of public TESS Alert data from pipelines at the TESS Science Office and at the TESS Science Processing Operations Center.
We make use of Filtergraph, an online data visualization tool developed at Vanderbilt University through the Vanderbilt Initiative in Data-intensive Astrophysics (VIDA).
The research shown here acknowledges use of the Hypatia Catalog Database, an online compilation of stellar abundance data as described in \citet{Hinkel14}, which was supported by NASA's Nexus for Exoplanet System Science (NExSS) research coordination network and the Vanderbilt Initiative in Data-Intensive Astrophysics (VIDA).
This research has made use of the NASA Exoplanet Archive, which is operated by the California Institute of Technology, under contract with the National Aeronautics and Space Administration under the Exoplanet Exploration Program.  
This research made use of Astropy,\footnote{http://www.astropy.org} a community-developed core Python package for Astronomy \citep{astropy:2013, astropy:2018}.  
We also used data products from the Widefield Infrared Survey Explorer, which is a joint project of the University of California, Los Angeles; the Jet Propulsion Laboratory/California Institute of Technology, which is funded by the National Aeronautics and Space Administration; 
the Two Micron All Sky Survey, which is a joint project of the University of Massachusetts and the Infrared Processing and Analysis Center/California Institute of Technology, funded by the National Aeronautics and Space Administration and the National Science Foundation; 
and the European Space Agency (ESA) mission {\it Gaia} (\url{http://www.cosmos.esa.int/gaia}), processed by the {\it Gaia} Data Processing and Analysis Consortium (DPAC, \url{http://www.cosmos.esa.int/web/gaia/dpac/consortium}). Funding for the DPAC has been provided by national institutions, in particular the institutions participating in the {\it Gaia} Multilateral Agreement.  
We acknowledge support for the KELT project through the Vanderbilt Initiative in Data-intensive Astrophysics, Ohio State University, and Lehigh University. 
Resources supporting this work were provided by the NASA High-End Computing (HEC) Program through the NASA Advanced Supercomputing (NAS) Division at Ames Research Center for the production of the SPOC data products.

%%%%%%%%%%%%%%%%%%%%%%%%%%%%%%%%%%%%%%%%%%%%%%%%%%%%%

\bibliographystyle{aasjournal}
\bibliography{references}

\end{document}